\documentclass{aa}
\usepackage{epsfig}
\usepackage{graphicx}
\def\araa{ARA\&A}%
\def\apj{ApJ}%
\def\apjl{ApJ}%
\def\aap{A\&A}%
\def\nat{Nature}%
\def\sgr{SGR~1806$-$20}
\def\lbv{LBV~1806$-$20}

\def\gc{G10.0$-$0.3}

\newcommand{\hi}{\mbox{H{\sc~i}}}
\newcommand{\hii}{\mbox{H{\sc~ii}}}

\def\co{$^{12}$CO(J=1--0)}
\def\tco{$^{13}$CO(J=1--0)}
\def\kms{\mbox{km s$^{-1}$}}
\def\av{A$_{\rm v}$}

\begin{document}
   \title{The Connection between W31, SGR 1806-20, \& LBV 1806-20: Distance,
Extinction, and Structure }

\titlerunning{W31, SGR 1806-20, \& LBV 1806-20}
\authorrunning{Corbel \& Eikenberry}


   \author{St\'ephane Corbel\inst{1}
          \and
          Stephen S. Eikenberry\inst{2,}\thanks{Visiting astronomer, Cerro Tololo
                Inter-American Observatory, National Optical Astronomy Observatories,
                which are operated by the Association of Universities for Research in
                Astronomy, under contract with the National Science Foundation.}
          }

   \offprints{S. Corbel}

   \institute{Universit\'e Paris VII and Service d'Astrophysique (F\'ed\'eration APC), 
		CEA Saclay, F-91191 Gif sur Yvette, France \\
		\email{corbel@discovery.saclay.cea.fr} 
	\and	
		Department of Astronomy, University of Florida, 
		211 Bryant Space Science Center, Gainesville, FL 32611, USA \\
	        \email{eiken@astro.ufl.edu}
	}
		
   \date{Received: ; Accepted: }

   \abstract{
We present new millimeter and infrared spectroscopic
observations towards the radio nebula G10.0$-$0.3, which is powered by
the wind of the Luminous Blue Variable star \lbv \, also closely
associated with the soft gamma-ray repeater \sgr, and believed
to be located in the giant Galactic \hii\ complex W31.  Based on
observations of $CO$ emission lines and $NH_3$ absorption features
from molecular clouds along the line of sight to G10.0$-$0.3, as well as
the radial velocity and optical extinction of the star powering the
nebula, we determine its distance to be 15.1$^{+1.8}_{-1.3}$ kpc in
agreement with Corbel et al. (1997).  In addition, this strengthens 
the association of \sgr\ with a massive molecular cloud at the
same distance. All soft gamma-ray repeaters  with precise location are now found to be 
associated with a site of massive star formation or molecular cloud. We also show that W31 consists
of at least two distinct components along the line of sight.  We
suggest that G10.2$-$0.3 and G10.6$-$0.4 are located on the $-$30
\kms\ spiral arm at a distance from the Sun of 4.5 $\pm$ 0.6 kpc and
that G10.3$-$0.1 may be associated with a massive molecular cloud at
the same distance as the LBV star, i.e. 15.1$^{+1.8}_{-1.3}$ kpc, 
 implying that W31 could be decomposed into two components along the line of sight.

	
   }

    \maketitle
%

\section{Introduction}
\begin{figure*}[hbt!]
\centerline{\epsfig{figure=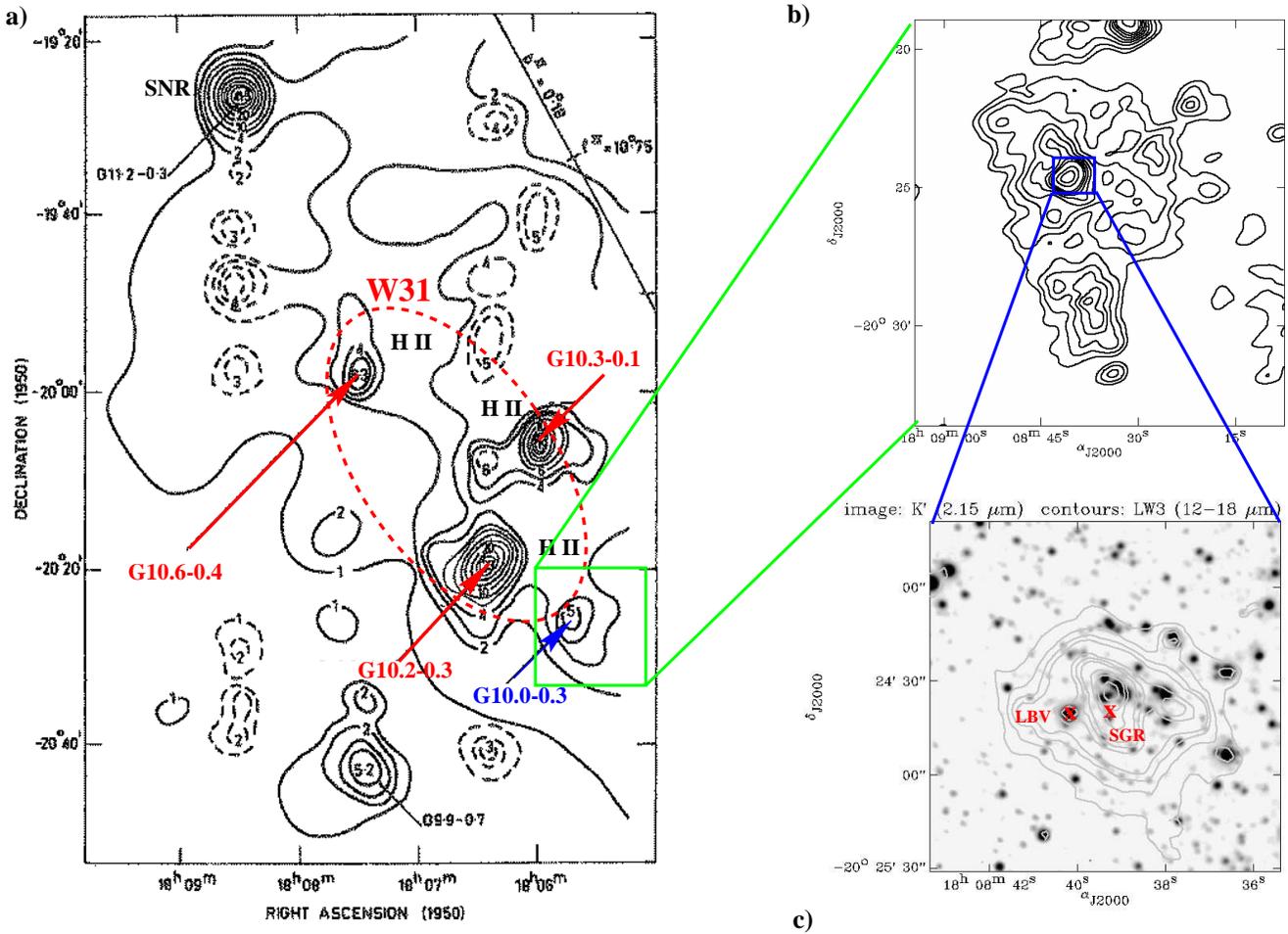, width=18cm}}
\caption{ Representation of the location of each object introduced
in this paper and their relative positions on the plane of the
sky. {\bf a)} Large radio map at 408 MHz of the field of W31, adapted from
Shaver \& Goss (1970). The location of the main components of W31 are
labeled, as well as the radio nebula G10.0$-$0.3 (produced by the wind
of \lbv ). {\bf b)} Radio image of the radio nebula G10.0$-$0.3 at 1.4 GHz
from Kulkarni et al. (1994). {\bf c)} K' (2.15 $\mu$m) near-infrared image
($\sim$ 2 \arcmin\ $\times$ 2 \arcmin) of the core of G10.0$-$0.3. The
location of \lbv\ and \sgr\ are indicated by a cross. The ISOCAM LW3
(12-18 $\mu$m) contours are superposed on this image in order to
highlight the spatial extend of the cluster of massive stars around
\sgr\ (adapted from Fuchs et al. 1999).  }
\end{figure*}

        W31 is one of the largest \hii\ complexes in the Galaxy, with
intense star-forming regions that have been observed from radio to
near-infrared wavelengths (e.g. Ghosh et al. 1989; Blum et al. 2001;
Kim \& Koo 2002). At low spatial resolution, W31 appears as three main
extended \hii\ regions: G10.2$-$0.3, G10.3$-$0.1 and G10.6$-$0.4
(Shaver \& Goss 1970) (Fig. 1).  The radio nebula G10.0$-$0.3 lies
within W31 on the plane of the sky (Fig. 1), and has drawn
considerable attention in recent years due to the intriguing objects
nearby.  At one time, G10.0$-$0.3 was suggested to be a plerionic
supernova remnant powered by a rare soft gamma-ray repeater, 
SGR~1806$-$20 (Kulkarni \& Frail 1993; Kouveliotou et al. 1998).  
SGR~1806$-$20, in turn, was
thought to be associated with an almost equally rare luminous blue
variable (LBV) star (van Kerkwijk et al. 1995) which lies at the
time-variable (in both flux and morphology) core of this nebula 
(Vasisht, Frail, \& Kulkarni 1995).
However, the revised Inter-Planetary Network (IPN) localization of SGR
1806$-$20 provides a position inconsistent (see Fig. 1) with that of
the LBV star and radio core of G10.0$-$0.3 (Hurley et al. 1999),
though the LBV position is consistent with the radio core within the
uncertainties ($\sim 2 \arcsec$).  Recent {\it Chandra} and infrared
observations confirm that the SGR lies $\sim 12 \arcsec$ away from the
LBV and radio core (Eikenberry et al. 2001; Kaplan et al. 2002).
Furthermore, Gaensler et al. (2001) argue that G10.0$-$0.3 is not a
supernova remnant at all, but is rather powered by the tremendous wind
of the LBV star at its core.  Infrared observations of the field of
SGR 1806$-$20 reveal that the LBV star is not alone, but appears to be
part of a cluster of embedded, hot, luminous stars (\cite{fuc99}), and
the IPN position for SGR 1806$-$20 is consistent with membership in
the star cluster (Eikenberry et al. 2001).  Given this somewhat confusing
history, we take a moment to summarize our current understanding of
G10.0-0.3: 1. G10.0-0.3 is a radio nebula (NOT supernova remnant) with
emission powered by the LBV star spatially coincident with its core.
2.  The LBV star is part of a cluster of luminous stars embedded in a
molecular cloud.  3. \sgr \ is likely another member of this cluster of
stars, and is spatially distinct from the LBV star. We plot in Fig. 1
the various objects (and their relative location on the plane of the
sky) that we will discuss in this paper.

       Because of the proximity of these unusual objects, the
distance to them can provide significant insight into their physical
properties, giving this measurement particular importance.  Corbel et
al. (1997) proposed a distance estimate based on observations of
molecular clouds along the line of sight.  They used $CO$ spectroscopy
to estimate the hydrogen column density towards G10.0$-$0.3, and from
that an absorption column density as a function of distance.  Taking
the measured X-ray absorption towards SGR 1806$-$20 and an estimate
of the optical extinction to the LBV star, they concluded that
G10.0$-$0.3, \sgr\ and W31 lie $14.5 \pm 1.4$ kpc from the Sun.

\begin{figure*}[hbt]
\centerline{\epsfig{figure=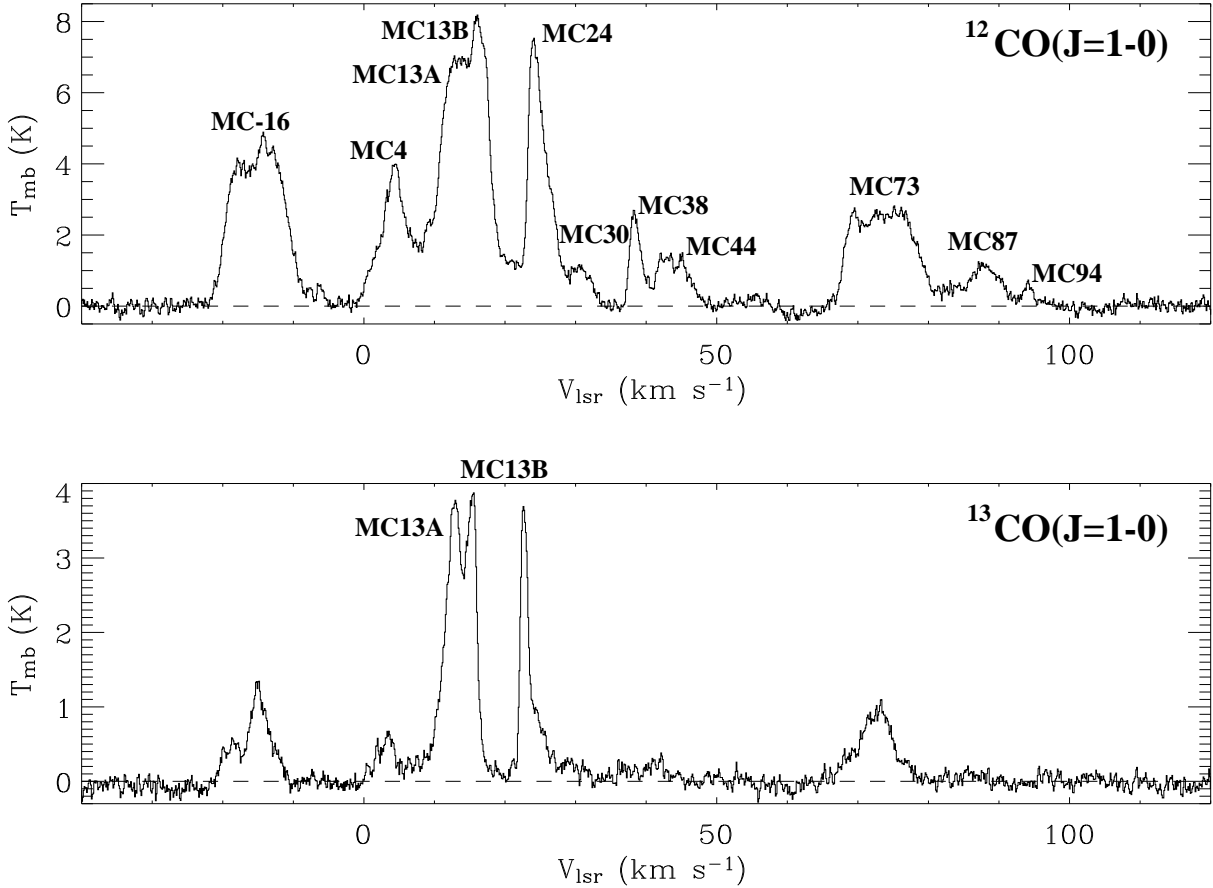,width=18cm}}
\caption{\co\ (top) and \tco\ (bottom) spectrum in the direction
of \lbv \ and \sgr.  Antenna temperatures have been converted into main
beam brightness temperatures. Each molecular cloud is labeled by
its name.
}
\end{figure*}

        However, this situation seemed complicated by newer infrared
stellar spectroscopy by Blum et al. (2001) and radio/millimeter
observations by Kim \& Koo (2002).  Blum et al. (2001) present
infrared spectra of members of a star cluster in the \hii\ region
G10.2$-$0.3, also within W31 on the plane of the sky.  Based on the
spectra, they derive spectral/luminosity classes and extinctions for
the stars, which, combined with infrared photometry, place them and
G10.2$-$0.3 unambiguously at a distance of $d \simeq 3.4$ kpc.
However, the extinction towards G10.2$-$0.3 is much smaller ($\Delta
A_V \approx 15$ mag) than the extinction towards \lbv \ and the star
cluster close to \sgr\ (Eikenberry et al. 2001) and the correlated X-ray
absorption towards \sgr\ (Mereghetti et al. 2000).  Furthermore, 21-cm continuum
and several molecular line maps of the region (Kim \& Koo 2002) show that
G10.0$-$0.3 is rather separated from the primary components of W31
(G10.2$-$0.3 and G10.3$-$0.1).  Thus, as of this writing, there is no
definitive physical linkage between \gc\ and the major components of
W31.  Together, these imply that G10.0$-$0.3 may lie at a different
distance along the line of sight than G10.2$-$0.3.  It is this new,
more-complicated situation which motivates us to reconsider the
distance to G10.0$-$0.3 (and by extension \lbv\ and \sgr ) in the
light of previous work by Corbel et al. (1997) and Blum et al. (2001)
as well as new observations.

   In this paper, we present newer, higher velocity -resolution
$CO$ spectroscopy towards \lbv \ and its associated radio nebula \gc,
and also towards 2 (out of 3) of the brightest \hii\ regions of W31.
We also present the $NH_3$ absorption spectrum originally mentioned in
Corbel et al. (1997).  We then add higher-resolution infrared
spectroscopic observations of the LBV star at the center of
G10.0$-$0.3.  In Section 2, we present the observations and data
reduction as well as the main results from our observations.  In
Section 3, we use these data to derive a robust distance estimate for
\lbv \ and \gc. We then discuss in some detail the implications of this
distance for \sgr \ and \lbv , as well as the structure of W31 as a whole .  Finally, in
Section 4 we present our conclusions.

\section{Observations}

{
\begin{table*}[hbt!]
\caption{Derived parameters for each of the molecular clouds along the line of sight of \lbv \ and \sgr,
using the \co\ spectra.}
\label{tab_av_1806}
\begin{center}
\begin{minipage}{17cm}
\renewcommand{\thempfootnote}{\alph{mpfootnote}}
\renewcommand{\thefootnote}{\alph{footnote}}
\renewcommand{\footnoterule}{}
\begin{tabular}{l|cccccc}
\hline
\hline

Name  &  W(CO*)\footnote{W(CO*) is the integrated line area of the cloud. The error on W(CO*) itself is negligeable compared 
to the 20\% uncertainty we assume to take into account the uncertainties in the various
conversion factors for $A_V$.}   & V$_{lsr}$\footnote{In the text, we use a systematic error of $\pm$ 10 \kms\
on the velocity of the cloud in order to derive a robust distance range ($>$ 2$\sigma$) for each of the molecular clouds.}  
                            &  A$_v$    &  Near Distance$^b$ & Far distance$^b$ & Distance$^b$ \\
      & (K \kms)  &  (\kms) &  (magnitude)  &       (kpc)   &   (kpc)     &   (kpc)  \\
\hline

MC--16  & 28.4  & --14.9 & 8.6 $\pm$ 1.7 &   n.a.     &   24.9    &  4.5    \\
MC4     & 12.7  &    4.3 & 3.9 $\pm$ 0.8 &    0.2      &  16.6     &  0.2    \\
MC13A & 32.8  &   12.7 & 10.0$\pm$ 2.0 &   1.7      &  15.1    &   15.1  \\
MC13B & 9.8   &   16.7 & 3.0 $\pm$ 0.6 &   2.2      &  14.5    &   4.5   \\
MC24    & 18.3  &   23.6 & 5.6 $\pm$ 1.1 &    3.0     &   13.8   &    3.0   \\
MC30    & 2.6   &   30.0 & 0.8 $\pm$ 0.2 &    3.5     &   13.2   &    3.5   \\
MC38    & 3.6   &   38.4 & 1.1 $\pm$ 0.2 &    4.2     &   12.6   &    4.2  \\
MC44    & 5.0   &   43.5 & 1.6 $\pm$ 0.3 &    4.5     &   12.3   &    4.5  \\
MC73    & 19.8  &   73.4 & 6.0 $\pm$ 1.2 &    5.7     &   11.0   &    ...\footnote{We did not
attempt to resolve the distance ambiguity for MC73, MC87 and MC94, since they are between
MC-16 and MC13A in either case.}  \\
MC87    & 4.5   &   87.9 & 1.4 $\pm$ 0.3 &    6.1     &   10.6   &    ...$^c$  \\
MC94    & 1.7   &   94.2 & 0.5 $\pm$ 0.1 &    6.2     &   10.5   &    ...$^c$   \\
\hline
\end{tabular}
\end{minipage}
\end{center}
\end{table*}
}

\subsection{$CO$ Spectra}

Following the work by Corbel et al. (1997), we obtained new millimeter
observations with the 15 m Swedish-ESO Submillimeter Telescope (SEST)
at La Silla, Chile, on 1998 August 27 and 1999 March 2.  We took
spectra at the position of LBV~1806-20 (same as in Corbel et al. 1997)
at the transitions $^{12}$CO(J=1--0) and $^{13}$CO(J=1--0) for a total
integration time of 5 minutes each (Fig. 2).  As the full-width at
half-maximum (FWHM) beamwidth of the SEST is 45\arcsec\ at $\sim$ 115 GHz,
the spectra toward \lbv \ also include the region of \sgr.
Additionally, at $^{12}$CO(J=1--0) we made spectral observations
towards two \hii\ regions of W31 (G10.2$-$0.3 and G10.3$-$0.1)
with an integration time of 2 minutes each.

We acquired the spectra in position-switching mode which consists of
switching (every one minute) between sources and an off-position free
of emission. We then averaged the spectra after baseline removal.  The
back end was an acousto-optical spectrometer with 2000 channels and a
frequency bandwidth of 86 MHz, giving a high velocity resolution of
0.11 km~s$^{-1}$ (to be compared with the velocity resolution of 2.3
km s$^{-1}$ in Corbel et al. 1997).  The receiver was calibrated with
the standard chopper-wheel method. Systems temperature during the
observations were typically in the range 200--350 K.  We converted all
CO spectra into main-beam brightness temperature ($\eta_{eff}$ = 0.7),
which is expressed in term of radial velocity in the Local Standard of
Rest (LSR).

We note that there is a distance ambiguity for any molecular cloud within the
solar circle at a given radial velocity (in the LSR). Indeed, at each
radial velocity it is possible to locate the cloud on the near side
(near distance) or on the far side (far distance) of the Galaxy.  We
assume an error of 10 \kms\ for the velocity-distance conversion using
the rotation curve of the Galaxy of Fich, Blitz \& Stark (1989) with
the standard rotation constants of R$_0$ = 8.5 kpc and $\Theta_0$ =
220 \kms. This is sufficiently large to cover any velocity deviation
($\sim$~4~\kms) with respect to the LSR frame of Galactic rotation
(Combes 1991). This implies that the error bars associated to the distances are
probably consistent with 2 $\sigma$ confidence levels. 

The $^{12}$CO spectra (Fig. 2) toward \lbv \ reveal a complex line of
sight, as already discussed in Corbel et al. (1997), with various
molecular clouds detected (see Table~1 for detailed informations). 
Due to the high velocity resolution, some
of the molecular clouds mentioned in Corbel et al. (1997) are now
split into two components; for simplicity the same notation is kept as
in Corbel et al. (1997). MC38 is decomposed into the component at
$\sim$ 38 km s$^{-1}$ and a new cloud at $\sim$ 44 km s$^{-1}$ (now
noted as MC44).  The resolution of the edge of the $^{12}$CO line of
MC24 also reveals a new cloud at 30 km s$^{-1}$ (MC30). MC87 is now
split into MC87 and MC94. However, the most interesting feature in the
new $^{12}$CO(J=1--0) spectrum toward \lbv \ is the splitting of
molecular cloud MC13 into two components. This is confirmed by the
$^{13}$CO(J=1--0) transition (an optically thin line, with narrower
FWHM), which clearly reveals that MC13 is resolved into two components
(Fig. 2): one at 13.0 $\pm$ 0.1 km s$^{-1}$ (hereafter MC13A) and one
at 15.5 $\pm$ 0.1 km s$^{-1}$ (hereafter MC13B).  (These velocities
are obtained if we naively fit the lines with Gaussian profiles).
This also implies that the shape of the $^{12}$CO(J=1--0) spectrum
around 13 km s$^{-1}$ is not due to opacity effects.  Fitting the
$^{12}$CO(J=1--0) spectra leads to velocities of 12.7 $\pm$ 0.1 km
s$^{-1}$ and 16.7 $\pm$ 0.1 km s$^{-1}$ for MC13A and MC13B
respectively.

According to Kim \& Koo (2001), the velocity of the recombination line
at the core of G10.2$-$0.3 is 16.4 $\pm$ 0.2 km s$^{-1}$, and it is
therefore apparent that G10.2$-$0.3 is associated with the molecular
cloud we labeled MC13B. We remind the reader that in addition to the
association in velocity, the map of the CO emission (Corbel et al.
1997) also points to an association of G10.2$-$0.3 with one of the
clouds in this velocity range. We note that the velocity of the
recombination line of the other major \hii \ region studied by Kim \&
Koo, G10.3$-$0.1, is 7.7 $\pm$ 0.5 km s$^{-1}$, with a range from 7.7
to 11.9 \kms (Kim \& Koo 2001), which is more consistent with the
velocity of MC13A.

\subsection{$\mathbf{NH_3}$ Spectrum}

In addition to the above CO observations, we also include an NH$_3$
absorption spectrum (Fig. 3) towards the radio nebula G10.0--0.3
(produced by the wind of \lbv) at the same position as the CO
observations. This spectrum was originally mentioned, but not shown,
in Corbel et al. (1997) as a ``Note added in manuscript''. As it is
important for this work, we include it here. The NH$_3$ observations
(total integration time of 30 minutes) were performed at the frequency
of 23,694.49 MHz (velocity resolution of 0.5 km s$^{-1}$) with the
NASA Deep Space Network Goldstone 34-m antenna located in California,
USA, and giving a beamwidth of 1.6\arcmin.
As the radio continuum against which the NH$_3$ absorption
spectrum is measured is produced by the radio nebula G10.0--0.3, any
absorption feature results from gas located in front of \lbv.

\begin{figure}[hbt]
\centerline{\epsfig{figure=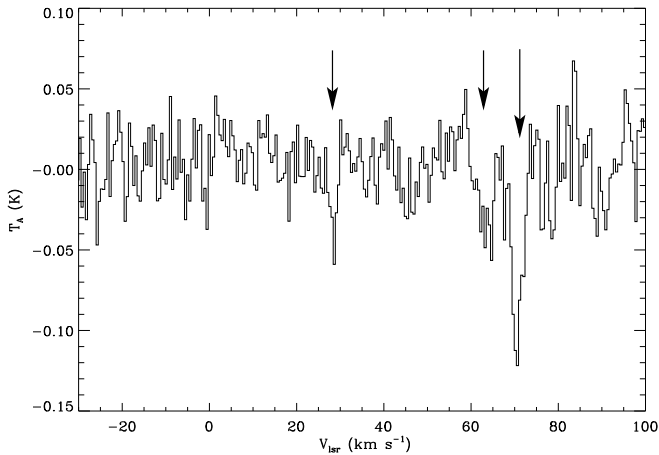, width=9cm}}
\caption{ NH$_3$ absorption spectrum measured against the radio
nebula G10.0$-$0.3 produced by the wind of \lbv. Arrows indicate the
main ($\sim 9 \sigma$) absorption feature at 71 \kms\ as well as the
possible features at 62 \kms\ ($\sim 4 \sigma$) and 29 \kms\ ($\sim 2 \sigma$)} .
\end{figure}

The most striking feature from this spectrum is the absorption line
(detected at a significance level of 8.6 $\sigma$) at a velocity of
70.9 $\pm$ 0.5 km s$^{-1}$, in full agreement with the $CO$ velocity
of the molecular cloud MC73 that is detected along the line of sight
to \lbv \ (see Table 1). This detection thus demonstrates that the
molecular cloud MC73 is located in front of \lbv. 
The absorption line ($4.2 \sigma$) at $\sim$ 62 \kms\  may be related to an \hi\ shell
around MC73 or to a different \hi\ cloud (cf Fig. 1 in Corbel et al. 1997)
in front of G10.0$-$0.3. A possible weaker
absorption feature ($2.2 \sigma$) might also be present at 28.6 km s$^{-1}$, 
which would be in agreement with the velocity of the MC30
cloud and would indicate that MC30 may also lie front of \lbv.

\subsection{IR Spectra}

        We used the Ohio State InfraRed Imaging Spectrograph (OSIRIS)
instrument (Depoy et al. 1993) and f/14 tip-tilt secondary on the
Cerro-Tololo Inter-American Observatory (CTIO) 4-meter telescope on
July 5, 2001 to obtain moderate resolution ($R = 3000$ for 2 pixels)
spectra of LBV 1806-20 in the J, H, and K bands ($1-2.4 \mu$m).  We
present details of these observations and their reduction elsewhere
(Eikenberry et al. 2003), and present a reduced K-band spectrum in Fig. 4.

\begin{figure}[hbt]
\centerline{\epsfig{figure=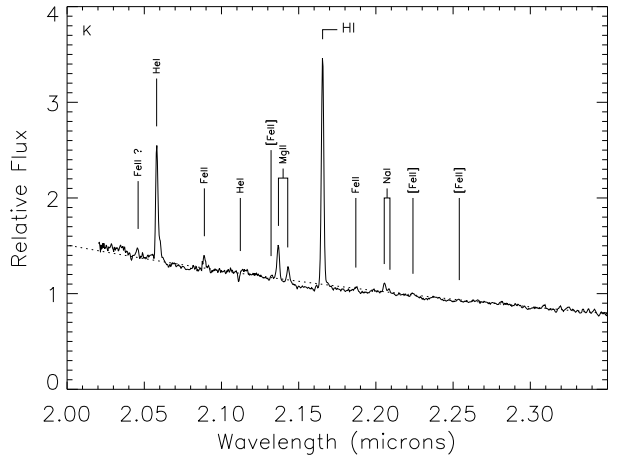,width=9.5cm}}
\caption{ Near-infrared spectrum of LBV 1806-20 in the K band,
de-reddened with $A_V = 29$ mag, following the reddening law of
Rieke \& Lebofsky (1985).  The dotted line indicates the spectral shape
of a blackbody with $T > 12000$ K.}
\end{figure}

        The colors of LBV 1806-20 (van Kerkwijk et al. 1995) and its
spectral continuum shape allow us to estimate the extinction towards
LBV 1806-20.  For such a hot star (as indicated by the HeI $2.112
\mu$m absorption feature -- Van Kerkwijk et al. 1995; Eikenberry et al. 2003), the intrinsic
$J-K$ color is nearly neutral, and the observed red color of $J-K =
5.0 \ \pm 0.15$ mag corresponds to an extinction of $A_V = 28 \pm 2$
mag (assuming the Rieke-Lebofsky reddening law, Rieke \& Lebofsky
1985), matching the estimates based on CO observations (Corbel et al. 1997).
(While a hypothetical near-infrared excess from LBV 1806-20 would
alter these conclusions, Eikenberry et al. (2003) show that this is
not present, based on the spectral continuum shape across JHK bands,
and the close match in $J-K$ color between \lbv \ and other cluster
stars).  In addition, the H and K bands are in the Rayleigh-Jeans
portion of the blackbody emission curve (the reason for the neutral
colors noted above).  Thus, we can estimate the extinction towards LBV
1806-20 by de-reddening the spectra until the continuum shape matches
a Rayleigh-Jeans distribution.  In this way, we obtain estimates of
$A_V = 31 \pm 3$ mag from the H-band continuum and $A_V = 28 \pm 3$
mag from the K-band continuum, with uncertainties dominated by $\sim
10\%$ uncertainty in the spectrograph response shape over a given
waveband.  Combining these with the extinction estimate from the $J-K$
color above, we adopt a final estimate for the extinction of $A_V = 29
\pm 2$ mag towards LBV 1806-20.

From the emission lines, we can also measure a radial velocity
for LBV 1806-20.  We selected the $\rm Br \gamma$ line as a velocity
fiducial, as it is the strongest line detection in the spectrum, and
appears to be relatively free from contamination due to blending with
other strong lines.  We fit a Gaussian profile to this line, finding
no significant residuals, and a centroid shifted from the atmospheric
rest frame by $-3 \pm 20 \ {\rm km \ s^{-1}}$, where residuals in the
spectral wavelength solution from atmospheric OH emission lines
dominate the largely systematic uncertainty.  After correcting for the
Earth's barycentric motion and the Solar System barycenter motion
relative to the local standard of rest, we determine a radial velocity
for LBV 1806-20 of $v_{lsr} = 10 \pm 20 \ {\rm km \ s^{-1}}$.
Cross-checks of this velocity determination with several other strong
unblended lines in the spectra give consistent results for the
velocity of LBV 1806-20.  This velocity is important, as massive stars
such as LBVs are a kinematically ``cold'' population, which do not
generally deviate significantly in their velocities from their parent
molecular clouds.

\section{Results \& Discussion}

\subsection{The distance to \lbv \ and its associated radio nebula \gc\ }

\subsubsection{$NH_3$ Absorption}

A critical point for measuring the distance to \lbv \ is the detection
of the NH$_3$ absorption features (Fig. 3) against the radio continuum
(G10.0$-$0.3) produced by the LBV star. A firm lower limit on the
distance to \gc\ and \lbv \ is the near distance associated with the
absorption feature at 70.9 $\pm$ 0.5 \kms, i.e 5.7 $\pm$ 0.4 kpc.  We
can then combine the full velocity range ($-$10 to 30 \kms) for \lbv \
(section 2.3 above) and the fact that \lbv \ has to be located behind
MC73, together with the rotation curve of the Galaxy (Fich et
al. 1989) to constrain the distance range for \lbv \, without making
any further assumptions regarding other molecular clouds along the
line of sight.  This straightforward calculation unambiguously places
\lbv \ in the distance range 13.2--21.5 kpc.

\subsubsection{MC13A, MC13B, and G10.2$-$0.3}

This determination of the distance to G10.0$-$0.3, while unambiguous,
differs significantly from the distance determination for the nearby
(on the sky) stellar cluster G10.2-0.3, which Blum et al. (2001) place
at $3.4 \pm 0.3$ kpc.  Given the similar radial velocities of \lbv \
and G10.2$-$0.3 ($\sim 10-16 \ {\rm km \ s^{-1}}$), this seems
somewhat surprising. However, on the plane of the sky, they are
separated by 10.7\arcmin. Furthermore, the measured optical extinction
of \lbv \ (29 $\pm$ 2 mag., section 2.3) and G10.2$-$0.3 (15.5 $\pm$
1.7 mag., Blum et al. 2001) are clearly not in agreement.  This
indicates that these two objects cannot be located at the same
distance. G10.2$-$0.3, with a lower optical extinction, has to be
located in the foreground relative to \lbv. Furthermore, Lavine et
al. (2003) find that the stellar field surrounding \lbv \ includes two
distinct populations of stars -- one with $A_V \sim 30$ mag
(consistent with \lbv ) and another with $A_V \sim 15$ mag
(consistent with G10.2$-$0.3).  Fuchs et al. (1999) also find the same
bimodality in extinction, albeit with a much smaller sample.  The fact
that no stars are found with intermediate extinctions demonstrates
that these are distinct populations, rather than a single population
suffering from differential extinction across the field.  Thus, we are
forced to conclude that the stellar cluster of Blum et al. (2001) is a
distinct cluster in the foreground to the cluster containing \lbv \ .
Thus, discrepant distances for these two distinct clusters are not
surprising, and in fact should be expected.

In order to understand the fact that these distinct populations have
similar radial velocities, we need to take into account (as outlined
in section 2.1) that the molecular cloud labelled MC13 in Corbel et
al. (1997) is now known to consist of two distinct components that we
have called MC13A and MC13B. Thus, the line of sight towards these
objects is much more complex than previously thought, with two
molecular clouds with similar velocities, one containing each stellar
population. Based on the fact that MC13B is associated with
G10.2$-$0.3 (section 2.1; Kim \& Koo, 2002), and that the upper limit for the
distance to G10.2$-$0.3 (Blum et al., 2001) is much less than the 13.2 kpc lower limit for
\lbv , we conclude that the molecular cloud MC13B is
located in the foreground relative to \lbv .

We can further constrain the parent molecular cloud of \lbv \ by
investigating the complex of molecular clouds along the line of sight.
Absorption lines against the radio continuum of G10.2$-$0.3 (see
Corbel et al.  1997 and references therein) showed that MC4, MC24,
MC30, MC38, MC44 are located at their near distances.  Given the
ammonia absorption towards \lbv \ and its accompanying distance range
of 13-21 kpc, the only remaining molecular clouds along the line of
sight which could potentially be the site of \lbv \ are MC87, MC94,
and MC13A.  Of these, only MC13A has a velocity consistent with \lbv \
(Section 2.3), and we conclude that it is the parent molecular cloud
for \lbv, \sgr, and \gc.  This association and the distance range for
\lbv \ now lift the near/far distance ambiguity for MC13A, and we can
use the velocity of MC13A and the Galactic rotation curve to constrain
the distance for both MC13A and \lbv \ to be 15.1$^{+1.8}_{-1.3}$ kpc.

\subsubsection{Extinction \& Distance}

We can cross-check the above distance estimation, using the measured
extinction towards \lbv \ .  This extinction arises primarily within
the molecular clouds along the line of sight, and we can determine its
value for each cloud using the $CO$ line area.  In order to convert
the molecular emission spectra into equivalent optical extinction
along the line of sight, we follow the method outlined in Corbel et
al. (1999) or Chapuis \& Corbel (2003). We assume conservative errors of 20\% (to take into
account the uncertainties in the various conversion factors) for the
contribution in terms of optical extinction of each molecular cloud.
The discussion of Corbel et al. (1997) for the location of the various
molecular clouds still holds with these new observations -- we simply
need to consider the revised distance of G10.2$-$0.3 and the splitting
of MC13. Separating the contribution of each of the two MC13 clouds in
terms of line area is not a simple task. But if we fit both
$^{12}$CO(J=1--0) and $^{13}$CO(J=1--0) spectra with two Gaussian
lines in the velocity range 8-20 km s$^{-1}$, we find that MC13B and
MC13A contribute roughly 23\% and 77\% respectively in terms of line
area. These should not be taken as firm numbers, but rather as
indicative of the approximate relative contribution of each clouds.
The resulting parameters (velocity, integrated area, optical
extinction, distances) of the molecular clouds are presented in
Table~1.

If we take the contribution from molecular hydrogen in all of the
clouds up to MC13A, we reach a total extinction of 32.5 $\pm$ 2.6
mag. Taking the contribution from atomic hydrogen (Corbel et al. 1997)
implies that the total optical extinction up to MC13A is 37.5 $\pm$ 3.0
mag.  While the uncertainties are not trivially small, this is at
least consistent (at the $\sim 2 \sigma$ level) with the extinction of
29 $\pm$ 2 mag for \lbv.  Adding the contribution of MC13A
significantly increases this extinction to 47.5 $\pm$ 3.6 mag, implying
that \lbv \ is at or close to the near side of MC13A.  In Fig. 5a, we
present a map of the integrated \tco\ emission (based on the data of
Kim \& Koo 2002).  It shows that the position of \lbv, as well as
\sgr\ and the star cluster, is consistent with a location on the edge
of the cloud MC13A, as favoured by the measurement of the optical
extinction.  Thus, we find a straightforward explanation for the
distance of G10.0$-$0.3 and \lbv , consistent with all reported data,
is that the star is located in MC13A at a distance of
15.1$^{+1.8}_{-1.3}$ kpc.

\subsection{The Distance to G10.2$-$0.3  and  the Complex Structure of W31 }

\subsubsection{Distance and optical extinction of G10.2$-$0.3}

With the distance of G10.0$-$0.3 and \lbv \ clarified, we now turn to
the more general issues of the structure of W31, which we now
understand to have several discrete components at distinct distances.
In their study, Blum et al. (2001) performed near-infrared
spectroscopy and photometry of an embedded stellar cluster in
G10.2$-$0.3 (the other major \hii\ regions of W31 being G10.3$-$0.1
and G10.6$-$0.4). They derived a spectrophotometric distance for
G10.2$-$0.3 (and by extension for W31) by assuming either zero-age
main-sequence or dwarf luminosity class for the stars of the cluster,
obtaining distances of 3.1 $\pm$ 0.3 kpc and 3.7 $\pm$ 0.3 kpc
respectively. They derived an average optical extinction to
G10.2$-$0.3 of \av\ = 15.5 $\pm$ 1.7 mag (Blum et al. 2001). Clearly,
this distance range, as well as the optical extinction, are not in
agreement with our previous estimate for W31 (Corbel et al. 1997), in
which we associate W31 with the far kinematic distance of
MC13. However, as shown above, we now understand that MC13 and W31
itself are composed of multiple discrete components distributed along
the line of sight.  Thus, as noted above, the distance to G10.2$-$0.3,
MC13B, and their portion of W31 is a separate issue from G10.0$-$0.3,
\lbv , \sgr, MC13A, and their portion of W31.  Nevertheless, our
investigation of the latter can provide important insights into the
distance to the former.

First, as a cross-check on the relative placement of molecular clouds
along the line of sight in Table 1, we can compare the extinction of
the stars in G10.2$-$0.3 to that expected from the clouds along the
line of sight (as done above for \lbv ).  If we use the \co\ spectrum
towards \lbv , the total visual extinction due to the molecular
material located in front of MC13B (MC4, MC24, MC30, MC38, MC44, cf
Corbel et al. 1997) is 13.0 $\pm$ 1.5 mag.  Adding the contribution
from atomic hydrogen (as in Corbel et al. 1997) raises this number to
14.5 $\pm$ 1.8 mag, consistent with the above estimate of the optical
extinction to G10.2$-$0.3 at the $\sim 1 \sigma$ level.  This
re-confirms the relative placement of the clouds in Table 1.

\subsubsection{G10.2$-$0.3: a location on the $-$30 \kms\ spiral arm}

\begin{figure*}[ht]
\centerline{\epsfig{figure=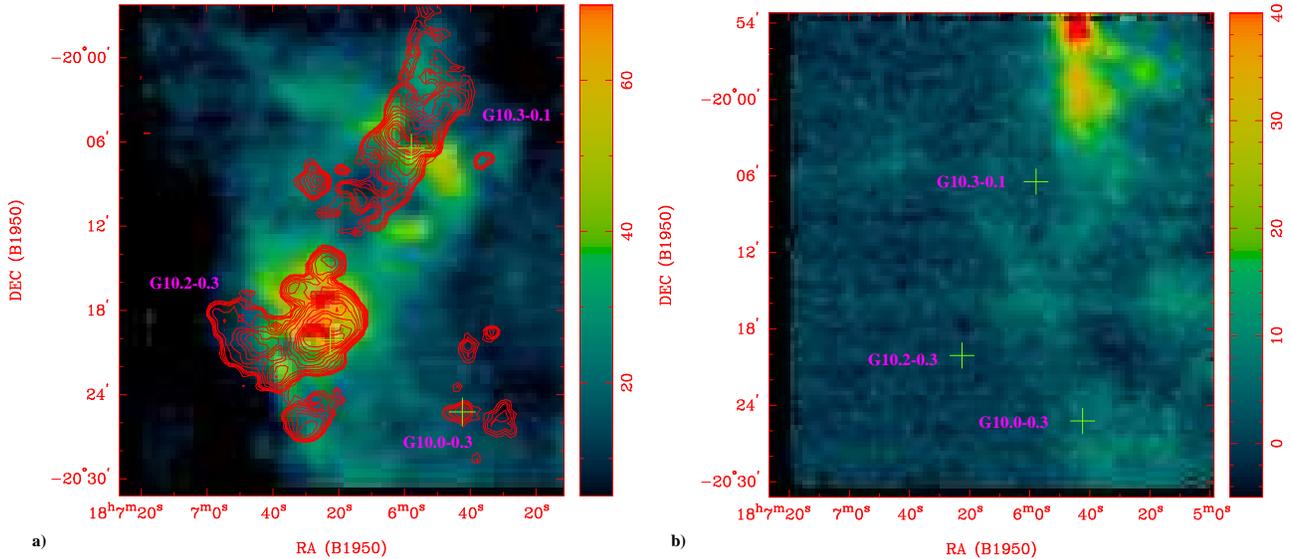,width=17cm}}
\caption{ {\bf a)} Map of the \tco\ emission integrated between the velocities 8  to 20 \kms, using the
data of Kim \& Koo (2002). {\bf b)} Map of the \tco\ emission integrated between the velocities 61  to 80 \kms. 
The location of G10.2$-$0.3, G10.3$-$0.1 and G10.0$-$0.3  are marked. The unit of the scale of 
the  map is in K \kms. The contours represent the 21 cm radio continuum
emission from this field of view (from Kim \& Koo 2002).
}
\end{figure*}
\begin{figure*}[hbt!]
\centerline{\epsfig{figure=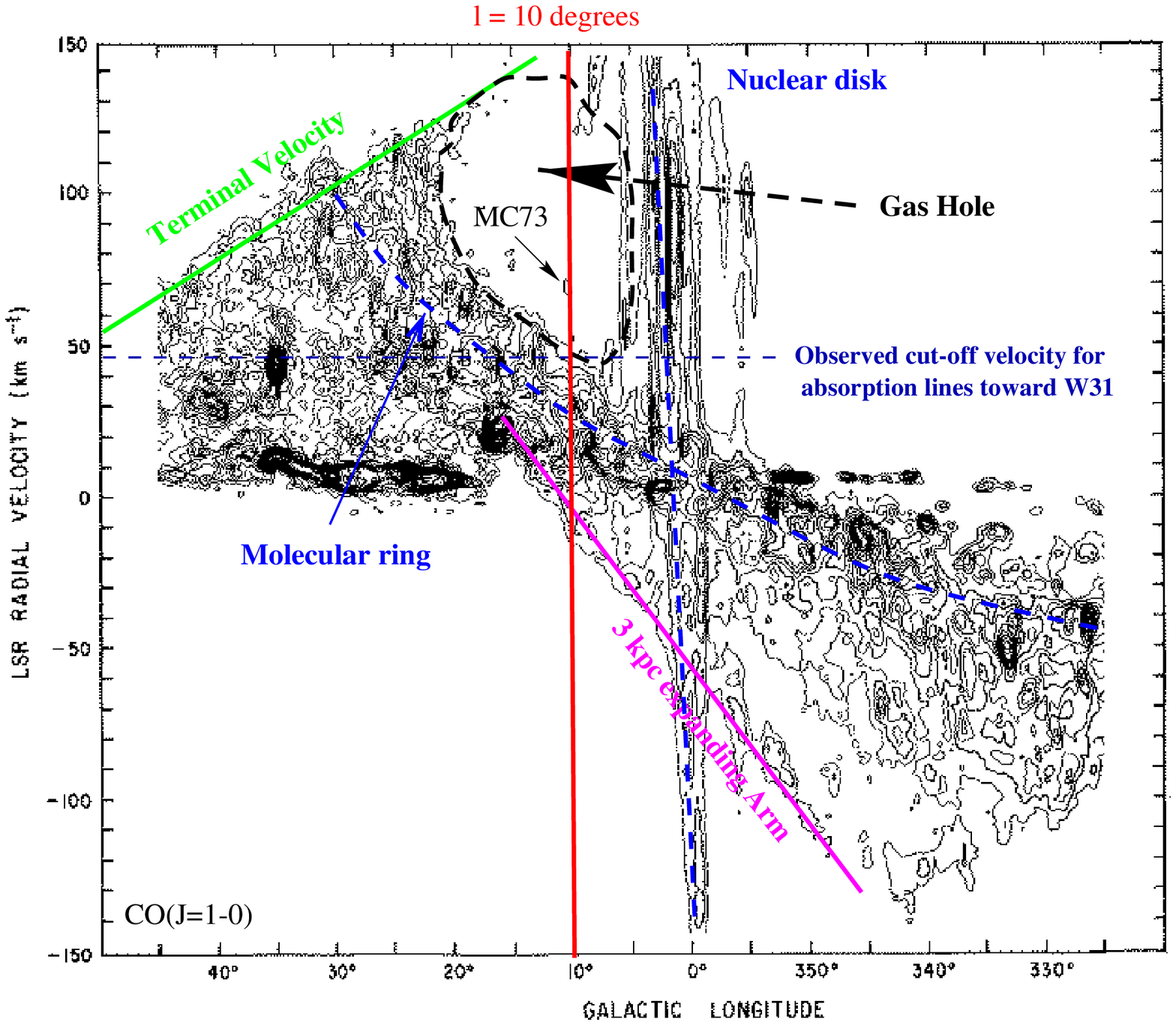,height=13cm}}
\caption{Longitude - velocity map of Galactic CO emission, Figure adapted from
Dame et al. (1987). Between Galactic longitude of $\sim$ 5\degr\ to 25\degr, a hole
is apparent in the gas distribution at velocities greater than $\sim$ 50 \kms.  We
note the location  of MC73 on the line of sight to G10.0$-$0.3. The locus point of
the 3 kpc expanding arm can also be well traced.}
\end{figure*}

Given this, an important question remains if we want to fully
understand the velocity field along this line of sight: what is the
distance to the parental molecular cloud (MC13B) of G10.2$-$0.3 ? A
distance range of 2.8-4.0 kpc (Blum et al. 2001) would imply a
velocity range of 22.0--38.0 \kms, which is at best marginally
consistent with the observed velocity of G10.2$-$0.3 (16.4 $\pm$ 0.2
\kms, Kim \& Koo 2001). Also, more importantly, absorption lines
against the radio continuum of G10.2$-$0.3 are observed up to 43 \kms\
(e.g. Wilson 1974, Greisen \& Lockman 1979, Kalberla et al. 1982),
which is very difficult to reconcile with G10.2$-$0.3. at the near
distance of 2.2$^{+1.0}_{-1.6}$ kpc associated with the velocity of
16.4 \kms.  Indeed, the cloud at the origin of the absorption at the
velocity of 43 \kms\ is located at a distance greater than 4.5 kpc
(see Table~1) if we assume circular motion. But we would like to note
that absorption at velocities greater than $\sim$ 50 \kms\ is not
expected at these Galactic longitudes, because of a hole (Fig. 6) in
the gas distribution (Dame et al. 1987; Corbel et al. 1997). The
drop-off of absorption at $\sim$ 43 \kms\ is just the sharp inner edge
of the molecular ring (Dame, Hartmann, \& Thaddeus 2001). Thus, the
lack of absorption should not be an argument for ruling out a possible
location at a far distance (e.g. Fish et al. (2003) for G10.6$-$0.4).

To overcome these difficulties, Blum et al. (2001), as well as other
authors (e.g. Wilson 1974, Kalberla et al. 1982), invoked the presence
of non-circular motion that could affect the velocity of the cloud (in
addition to the contribution of rotation around the Galactic Center,
hereafter GC).  Indeed, along the line of sight of W31 (and \lbv) is
found the 3-kpc expanding arm.  This arm can be described as a simple
rotating ring at a galactocentric radius of 3.4 kpc (assuming a
distance between the Sun and the GC of 8.5 kpc \footnote{Note that in
old literature, this arm was sometimes called the ``4-kpc expanding
arm'' due to a distance to the GC of 10.0 kpc.}) expanding out from
the GC with a velocity of 53 \kms\ (e.g. Bania 1980).

As discussed in detailed in Corbel et al. (1997), the 3-kpc expanding
arm is likely associated with the molecular cloud MC-16 and is very
likely not associated with MC13B and G10.2$-$0.3. This is confirmed if
we look at the radial velocity profile of the 3-kpc expanding arm as a
function of the Galactic longitude based on \hi\ observations (see
Figure 4 of Menon \& Ciotti, 1970). This clearly show that a velocity
of $\sim$ $-$15 \kms\ (as for MC-16) should be expected for this
feature.

So, in order reconcile the fact that G10.2$-$0.3 is located on this
side of the Galaxy (Blum et al. 2001) and the fact that absorption
lines are observed up to a velocity of 43 \kms, we surmise that there
might be another feature with non-circular motion on this side of the
Galaxy. In fact, there is another arm, the $-$30 \kms\ spiral arm
(unfortunately sometimes called the 4 kpc arm), that is also expanding
from the GC at a galactocentric radius of 4 kpc (Menon \& Ciotti 1970;
Greaves \& Williams 1994).  Its expansion velocity measured at a
Galactic longitude of 0\degr\ is $-$30 \kms\ (Menon \& Ciotti 1970;
Liszt et al. 1977; Linke, Stark, \& Frerking 1981; Greaves \& Williams
1994; Sandqvist et al. 2003).  This feature was originally detected in
1967 (Kerr \& Vallak 1967), but we would like to point that no
molecular counterpart has been associated with it. Indeed, unlike the
3 kpc expanding arm, the $-$30 \kms\ spiral arm can not be traced on
the longitude$-$velocity CO map (Fig. 6), possibly due to its
proximity with the molecular ring (Dame et al. 2001).  The
extrapolation of its radial velocity profile in Figure 1 of Menon \&
Ciotti (1970) to a longitude of 10\degr\ is consistent with the
velocity of MC13B.  We therefore conclude that G10.2$-$0.3 and MC13B
are located on the $-$30 \kms\ spiral arm.  In that case, there is no
problem with the fact that absorption lines against the radio
continuum of G10.2$-$0.3 are detected up to 43 \kms, as they would
originate in MC44. We note that the $-$30 \kms\ spiral arm has to be
closer to the Sun than the 3-kpc expanding arm with a separation of
$\sim$ 0.5 kpc (Menon \& Ciotti 1970).  In any case, the distance to
the $-$30 \kms\ spiral arm is set by the maximum velocity of the
absorption lines, i.e. 4.5 $\pm$ 0.6 kpc, which is completely
consistent with the spectrophotometric distance of 3.4 $\pm$ 0.6 kpc
for G10.2$-$0.3 (Blum et al. 2001).  We note that at a Galactic
longitude of 0\degr, the 3 kpc expanding arm and the $-$ 30 \kms\
spiral arm have a velocity separation of $\sim$ 23 \kms, which is
almost the difference in velocity between MC-16 and MC13B. This
therefore strengthens our association of the 3-kpc expanding arm with
MC-16 and the $-$30 \kms\ spiral arm with MC13B.  We also note that
these two arms could be related to the presence of a bar at the GC
(e.g. Blitz \& Spergel 1991).

\subsubsection{The distance to G10.3$-$0.1 and G10.6$-$0.4}

So if G10.2$-$0.3 is at a closer distance, are the other major
components of W31 at the same distance ?  In their detailed study of
W31, Kim \& Koo (2002) performed a \tco\ map of G10.2$-$0.3 and
G10.3$-$0.1 (see their Figures 4, 5 and 9). Their peak CO maps over
the velocity range 0--22 \kms\ shows two main components (one centered
on each \hii\ region) that could be interpreted as two separate
molecular clouds (see also Fig. 5a with the integrated \tco\ map). 
But based on our new CO results, it might be possible that the
southern part is associated with MC13B as G10.2$-$0.3, and the
northern part with MC13A and G10.3$-$0.1 (Fig. 5a). If this is the
case, it would imply that W31 could be decomposed into several
components, with G10.3$-$0.1 located at the kinematic distance
associated with MC13A, i.e. 15.1$^{+1.8}_{-1.3}$ kpc.

\begin{figure}[b]
\centerline{\epsfig{figure=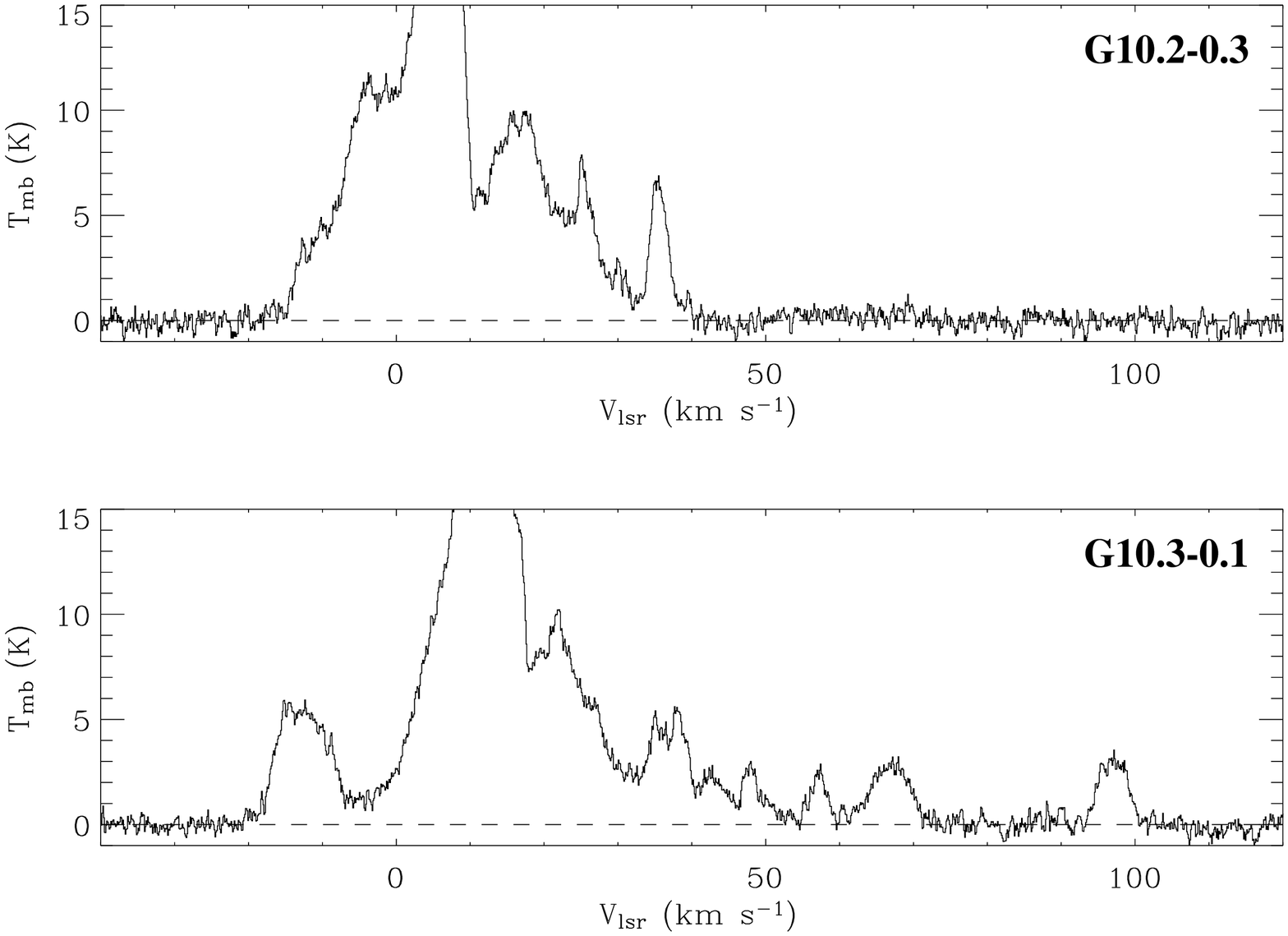,width=9cm}}
\caption{ \co\ spectrum in the direction  of the \hii\ regions G10.2$-$0.3 (top)  and
G10.3$-$0.1 (bottom).  Antenna temperature have been converted into main
beam brightness temperature.  The scale has been adjusted in order to highlight the
emission above 40 \kms.
}
\end{figure}

\begin{figure*}[bt!]
\centerline{\epsfig{figure=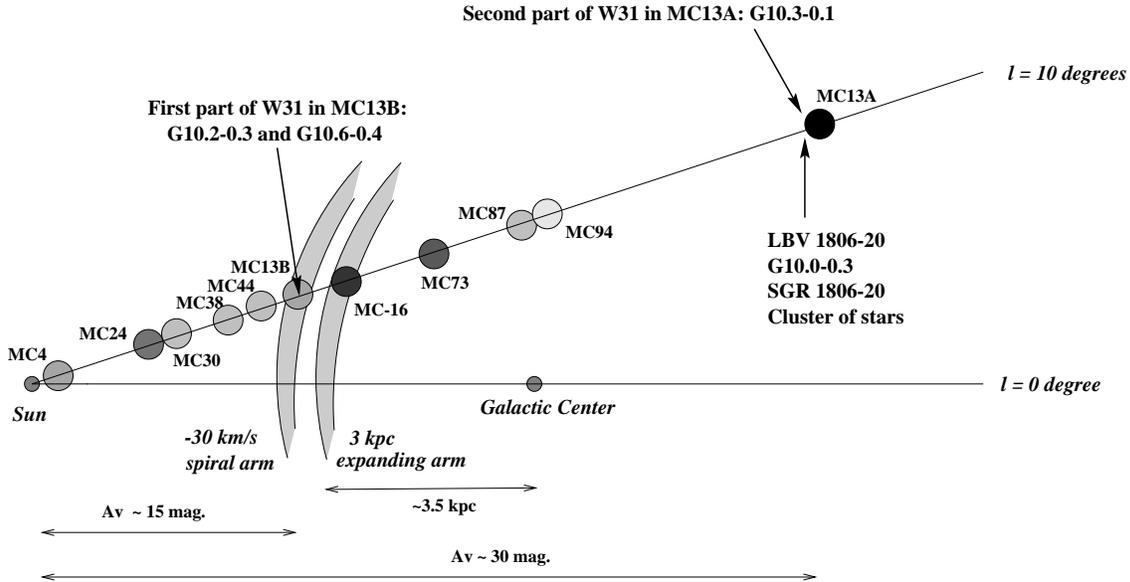,width=15cm}}
\caption{Schematic diagram of the molecular clouds on the line of sight to W31, \lbv\ and \sgr.}
\end{figure*}

Fig. 7 shows the $^{12}$CO spectrum along the line of sight to these
two \hii\ regions. These profiles are very different, especially at
velocities above $\sim$ 40 \kms, and the profile of G10.3$-$0.1 is
very similar to the one of \lbv\ (the additional CO emission above 40 \kms\ may 
also be related to the variation in galactic latitude of these sources). 
So it is not unlikely that this \hii\
region, which has a recombination line at 7.7 $\pm$ 0.5 \kms, could be
associated with MC13A. As noted above (section 2.1), the \tco\ map of
this part of W31 by Kim \& Koo (2002) could be interpreted as being
due to the presence of two separate molecular clouds. In that case,
one should wonder why no absorption line is detected, as in the case
of \lbv, at $\sim$ 71 \kms\ for G10.3$-$0.1 (Kalberla et al. 1982) ?
In Fig. 5b, we used the \tco\ of Kim \& Koo (2002) to illustrate the
spatial extent of MC73. We found almost no \tco\ emission in the
velocity range 61 to 80 \kms, which is consistent with the non
detection of absorption line above 50 \kms\ (Kalberla et al. 1982)
toward G10.3$-$0.1, even if it were associated with MC13A. However,
our \co\ spectrum towards G10.3$-$0.1 (Fig. 7) indicates a weak
contribution of MC73 along this line sight and suggest that the line
of sight of G10.3$-$0.1 might be close to the edge of MC73.  More
sensitive absorption measurement may be useful to further detect
absorption line due to MC73 in front of G10.3$-$0.1.
We note that Kalberla, Goss \& Wilson (1980) detected a weak absorption 
line at $\sim$ --15 \kms\ in front of G10.3$-$0.1, that is consistent 
with the velocity of the 3-kpc expanding Arm (Corbel et al. 1997), i.e. 
the molecular cloud we called MC-16 and the fact that G10.3$-$0.1 
could be associated with MC13A. We note, however, that a similar but weaker
feature, is also present in front of G10.2$-$0.3 (Kalberla et al. 1980).

In a recent study of absorption lines toward a large number of
\hii\ regions, Fish et al. (2003) deduced a scaling law in order to
estimate the distance of an \hii\ region.  They state that when there
is a ``large difference between far and near kinematic distances'' |as
in our case for both MC13A and MC13B, see Table 1| ``high accuracy can
be achieved by choosing the kinematic distance closer to 1.84
$|b|^{-1}$.''  In the case of G10.2$-$0.3, G10.3$-$0.1 and
G10.6$-$0.4, this expression would be equal to 5.3, 12.3 and 4.7 kpc
respectively with high accuracy \footnote{Note that this argument
cannot be applied to G10.0$-$0.3, simply because it is not an \hii\
region.}.  This again will argue for an association of G10.2$-$0.3 and
G10.6$-$0.4 with MC13B in the $-30$ \kms\ spiral arm at a distance of
4.5 $\pm$ 0.6 kpc and a location of G10.3$-$0.1 in MC13A at a distance
of 15.1$^{+1.8}_{-1.3}$ kpc.  But again, as illustrated in Fig. 3 of
Dame et al. (1987), between longitude 5\degr\ and 25\degr, we want to
stress that the lack of absorption at velocities greater than $\sim$
50 \kms\ is not an argument for the near distance due to the presence
of the gas hole (Fig. 6) at low Galactic longitudes (Corbel et
al. 1997). 

Finally, to close the debate on the structure of W31 (as a
location of G10.3$-$0.1 within the $-30$ \kms\ spiral arm can not be
ruled out at this time), a spectrophotometric study (with optical
extinction measurement), similar to what has been performed by Blum et
al. (2001) for G10.2$-$0.3, should also be performed for G10.3$-$0.1
and G10.6$-$0.4 in light of our new results.  A sketch of the line of
sight with the new location of each object introduced in this paper is
presented in Fig. 8.

\subsection{The distance to \sgr\ }

As we have seen before, \gc\ is associated with the wind of \lbv \ and
not \sgr\ and is not the remnant of a supernova (Gaensler et
al. 2001).  \sgr\ is separated from \lbv \ by an angular distance of
12 \arcsec\ (Hurley et al. 1999; Eikenberry et al. 2001; Kaplan et
al. 2002).  However, it still lies within the angular extent of the
embedded cluster (see Fig. 1 and Fuchs et al. 1999), 
and its X-ray absorption matches the IR extinction
towards the cluster members (Eikenberry et al., 2001), leading to the
conclusion that \sgr\ is a cluster member.  This lead to the conclusion
that all these objects, i.e. \sgr\ and the cluster of stars, have to
be located at the distance we have estimated for \lbv:
15.1$^{+1.8}_{-1.3}$ kpc. An interesting consequence of this work is
that all these massive stars are still located close to their parental
molecular cloud (MC13A), which is not surprising if we take into
account their short lifetimes and the expectation that such massive
stars are a kinematically ``cold'' population.  We would like to point
out that SGR~1900$+$14 is also associated with a cluster of massive
stars (Vrba et al. 2000) and that SGR~1627$-$41 also appears to lie at the edge
of a massive Giant Molecular Cloud (Corbel et al. 1999). 
The only known extragalactic soft gamma repeater (SGR~0526$-$66), which  is
associated with the SNR N49, also lies at the edge of a dense molecular cloud (Vancura et 
al. 1992; Banas et al. 1997). It now also means that all SGRs with precise location
(4 out 6) are associated with GMC and/or massive star cluster.
All of this probably points to a strong connection between massive stars 
and formation of SGRs by the way of Giant Molecular Clouds.

\section{Conclusions}

We have presented new millimeter and near-infrared observations of the
field of view surroundings the radio nebula G10.0$-$0.3 (produced by
the wind of \lbv) and the giant \hii\ complex W31. Based on these
observations combined with others in the literature, we reach the
following conclusions:

\begin{itemize}

\item Based on $NH_3$ absorption from MC73 and the velocity of \lbv \ ,
we unambiguously constrain the distance to G10.0$-$0.3 and \lbv \ to be
in the range of 13.2 - 21.5 kpc.

\item Combining this constraint with $CO$ observations of molecular
clouds along the line of sight, we further refine the distance
measurement to G10.0$-$0.3 and \lbv \ to be 15.1$^{+1.8}_{-1.3}$ kpc.

\item This distance estimate is confirmed by the consistency between
the measured extinction towards \lbv \ and the extinction from the
molecular clouds along the line of sight inferred from their $CO$
spectra.

\item Based on their distinct extinctions and the newly-resolved
parental molecular clouds (MC13A and MC13B), we conclude that the
stellar cluster in G10.2$-$0.3 lies in the foreground to the cluster
containing \lbv \ .  This shows that W31 consists of at least 2 discrete
components along the line of sight.

\item We suggest that G10.2$-$0.3 and G10.6$-$0.4 are located on the
  $-$30 \kms\ spiral arm at a distance from the Sun of 4.5 $\pm$ 0.6
  kpc and

\item We also suggest that G10.3$-$0.1 may be associated with a
  massive molecular cloud at the same distance as \lbv \
  (15.1$^{+1.8}_{-1.3}$ kpc).

\item We confirm that \sgr\ is located at a distance from the
Sun of 15.1$^{+1.8}_{-1.3}$ kpc and that it is associated with a
very massive molecular cloud. All SGRs with precise location
are associated with a site of massive star formation.

\end{itemize}


\begin{acknowledgements}
The authors would like to thank Tom Dame  and  Kee-Tae Kim for useful discussions and for
critical review of this manuscript. We also wish to thank Claude
Chapuis for conducting the 1998 SEST observations and stimulating
discussion and Bill Mahoney for carrying out the DSN observations.  We
thank R. Blum and the CTIO staff for their help in acquiring the IR
spectra.  We gratefully acknowledge Kee-Tae Kim and Bon-Chul Koo for
sharing their CO data of W31, as well as Yael Fuchs for her help with
Fig. 1. We also thank Aage Sandqvist for providing information on the
$-$30 \kms\ spiral arm.  SSE is supported in part by an NSF CAREER
award (AST-9983830).
\end{acknowledgements}

\clearpage
\newpage

\end{document}